\documentclass[aps,prb,reprint,superscriptaddress,showpacs]{revtex4-1}
\usepackage{graphicx}
\usepackage{subfigure}
\usepackage{dcolumn}
\usepackage{bm}
\usepackage{hyperref}
\usepackage{indentfirst}
\usepackage{braket}
\usepackage{amsmath}
\usepackage{amssymb}
\usepackage[normalem]{ulem}
\hypersetup{colorlinks=true, citecolor=blue, urlcolor=blue, linkcolor=blue}

\begin{document}

\title{Coexistence of non-Abelian chiral spin liquid and magnetic order in a spin-1 antiferromagnet}

\author{Yixuan Huang}
\affiliation{Department of Physics and Astronomy, California State University, Northridge, California 91330, USA}

\author{W. Zhu}
\email{zhuwei@westlake.edu.cn}
\affiliation{Key Laboratory for Quantum Materials of Zhejiang Province, School of Science, Westlake University, Hangzhou 310024, China}

\author{Shou-Shu Gong}
\affiliation{Department of Physics, Beihang University, Beijing 100191, China}

\author{Hong-Chen Jiang}
\affiliation{Stanford Institute for Materials and Energy Sciences, SLAC and Stanford University, Menlo Park, California 94025, USA}

\author{D. N. Sheng}
\email{donna.sheng1@csun.edu}
\affiliation{Department of Physics and Astronomy, California State University, Northridge, California 91330, USA}

\date{\today}

\begin{abstract}
We study the ground-state properties of a spin-1 Heisenberg model on a square lattice with the first- and second-nearest-neighbor antiferromagnetic couplings $J_1$ and $J_2$ and a three-spin scalar chirality term $J_\chi$. 
Using the density matrix renormalization group calculation, we map out a global phase diagram including various magnetic order phases and an emergent quantum spin liquid phase.  
The nature of the spin liquid is identified as a bosonic non-Abelian Moore-Read state from the fingerprint of the entanglement spectra and identification of a full set of topological sectors. 
We further unveil a stripe magnetic order coexisting with this spin liquid. 
Our results not only establish a rare example of non-Abelian spin liquids in simple spin systems but also demonstrate the coexistence of fractionalized excitations and magnetic order beyond mean-field descriptions.
\end{abstract}


\maketitle

\section{Introduction} 
Quantum spin liquids (QSLs) are  novel quantum states with long-range entanglement and emergent fractionalized excitations, which can avoid forming conventional magnetic order due to geometric frustration and quantum fluctuation~\cite{Anderson1973,SavaryBalentsReview,Zhou2017,Balents2010,Broholm2020}. 
The prominent realizations of QSLs are the chiral spin liquids (CSLs)~\cite{Kalmeyer1987}, which break time-reversal symmetry and have been established in some spin-$1/2$ systems~\cite{Bauer2014,He2014,Gong2014,Messio2012,Wietek2015,He2015c,Hu2015,Gong2015,Kumar2014,Hickey2016,Szasz2020,Chen2021}. 
These CSLs have fractional excitations following the Abelian anyon statistics, which are spin analog of the Laughlin state and bridge QSLs and the fractional quantum Hall effect~\cite{Wen1989,Wen1991,Wen1990a}. 
More interestingly, exactly soluble models host a new class of CSLs with non-Abelian quasi-particles~\cite{Kitaev2006,HYao2007} that have the potential to perform topological quantum computation~\cite{Nayak2008} and have stimulated an extensive search for CSLs in Kitaev materials with strong spin-orbit couplings~\cite{Hermanns2018}. 
Another natural system to search for non-Abelian CSLs is the frustrated spin-1 model~\cite{Greiter2009,scharfenberger2011non,Greiter2014,Glasser2015,Tobias2015}, which can be realized in both magnetic compounds~\cite{Nakatsuji2005,Cheng2011} and cold atom systems~\cite{Gorshkov2010}. 
However, unambiguous identification of non-Abelian CSLs in frustrated spin-1 systems is still rare so far~\cite{JYChen2018}.

While competing interactions and quantum fluctuations play important roles in forming a QSL, the interplay of different orders in the emergence of a QSL remains less understood. 
An interesting possibility is that a QSL coexists with a conventional order. 
So far, theoretical understandings of such an exotic coexistence have mainly been based on mean-field analyses~\cite{ZXLiu2010,Savary2012,Sedrakyan2015,Chern2019,samajdar2019enhanced}, and no example has been found with numerical calculation.
In some recent experiments, evidence of fractionalized excitations was found in materials with magnetically ordered ground states, such as the scattering continuum in the pyrochlore magnet Yb$_2$Ti$_2$O$_7$~\cite{Thompson2017}, which demands further theoretical understanding of the possible coexistence of QSL and magnetic order.

\begin{figure*}
\subfigure{\includegraphics[width=0.6\linewidth]{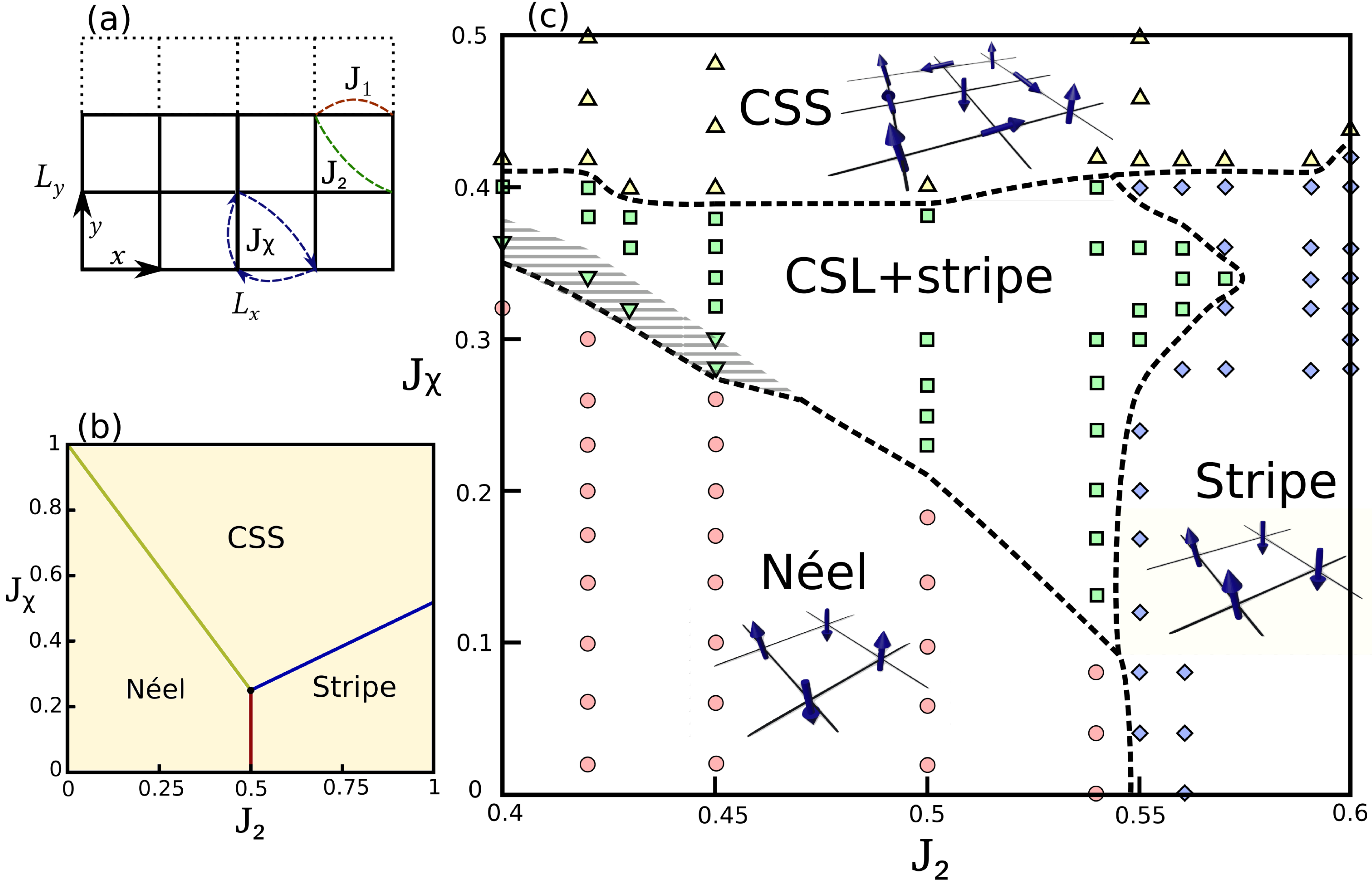}}
\subfigure{\includegraphics[width=0.39\linewidth]{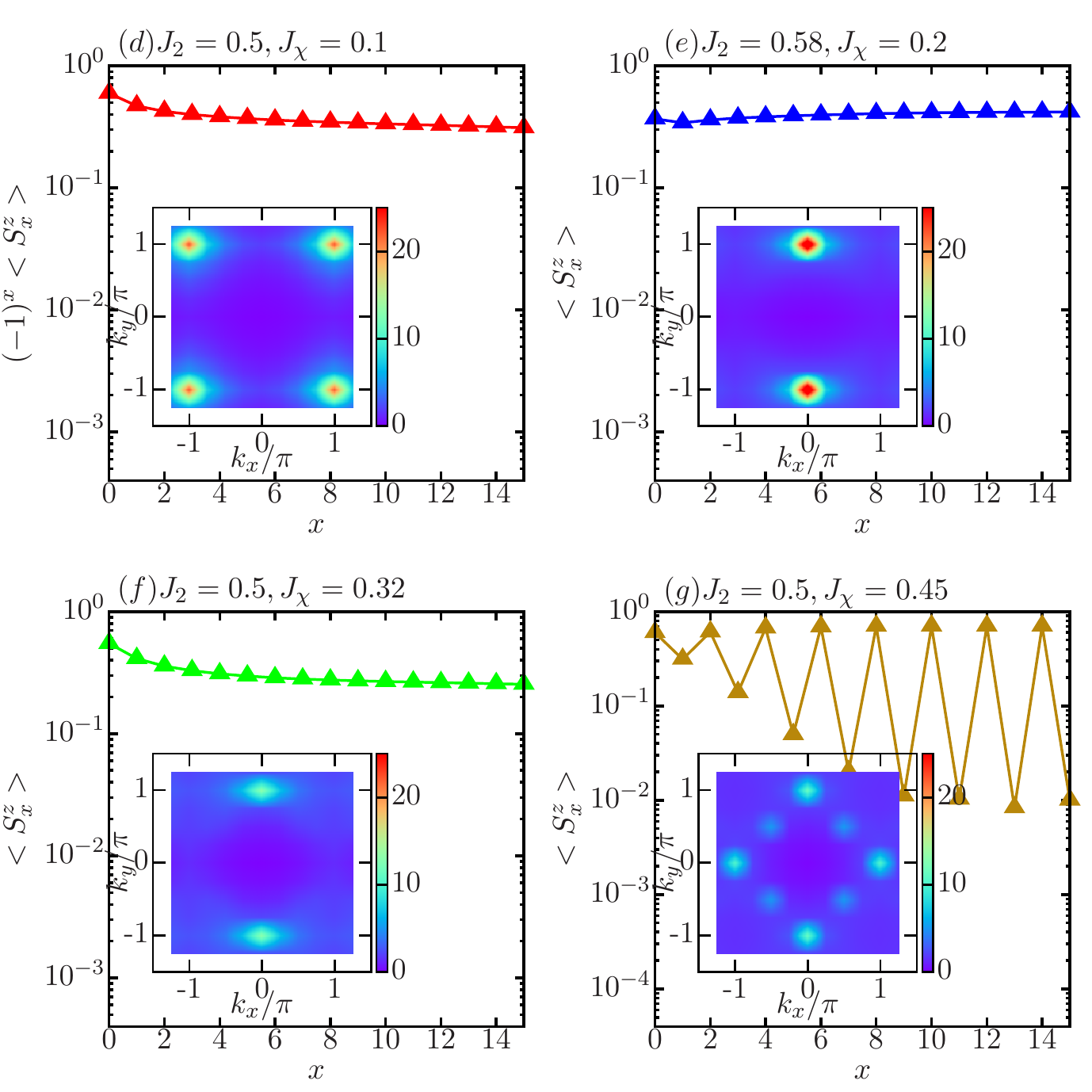}}
\caption{Model Hamiltonian and phase diagram. (a) The model and (b) classical phase diagram are shown on the left; the dashed lines in (a) indicate the periodic boundary conditions of the cylinder geometry. (c) depicts the quantum phase diagram of the spin-1 $J_{1}$-$J_{2}$-$J_\chi$ Heisenberg model with $0.4 < J_{2} < 0.6$ and $0 < J_{\chi} < 0.5$. The stripe order is suppressed in the shaded area, where the topological order remains robust.
The magnetic moments $\langle S^{z} \rangle $ are shown as a function of the distance $x$ away from the pinning field $h_{z}=0.1$ at the boundary for (d) N\'eel, (e) stripe, (f) CSL+stripe, and (g) chiral spin state. Smaller $h_{z}$ were also tested and give similar $\langle S^{z} \rangle $ in the bulk (see Appendix~\ref{supp_convergence}). The insets show the spin structure factors in the absence of a pinning field, which is defined as $S(\mathbf{k})=\frac{1}{N}\sum_{i,j}\left \langle \mathbf{S}_{i} \cdot \mathbf{S}_{j} \right \rangle e^{i\mathbf{k}\cdot (\mathbf{r}_{i}-\mathbf{r}_{j})}$. The results in (d)-(g) are obtained with $L_{y}=8$.}
\label{Fig_phase_diagram}
\end{figure*}

In this work, we study the roles of competing interactions in driving a QSL in a spin-1 square-lattice antiferromagnetic system with nearest-neighbor (NN) $J_1$ and next-nearest-neighbor (NNN) $J_2$ interactions and a three-spin scalar chiral coupling $J_\chi$. By means of an unbiased density matrix renormalization group (DMRG) approach, we map out a global phase diagram of the model, including a N\'eel magnetic state, a stripe magnetic state, a noncoplanar chiral spin state (CSS), and a CSL state surrounded by these conventional magnetic phases. 
We unambiguously identify this CSL as the non-Abelian Moore-Read state from the fingerprint of entanglement spectra (ESs) and the identification of the full topological degeneracy. 
Furthermore, we determine a  stripe magnetic order in this CSL phase, showing a coexistence of QSL and magnetic order.
Finally, the nature of the quantum phase transitions from the CSL to the neighbor phases is also addressed.
Our results show the presence of a robust non-Abelian CSL coexisting with the stripe magnetic order, and demonstrate a rare example of simple spin-1 systems resulting from the interplay between competing interactions and quantum fluctuations.

The rest of the paper is organized as follows: In Sec.~\ref{modelandmethod}, we introduce the $J_{1}$-$J_{2}$-$J_\chi$ model on a square lattice and describe the numerical method used in this work. In Sec.~\ref{phasediagram}, we summarize our main findings in the phase diagram including various magnetic order phases and an emergent CSL phase. We further show numerical evidence of the magnetic order phases in Sec.~\ref{magnetic}. In Sec.\ref{coexisting}, we identify the nature of the spin liquid as the non-Abelian CSL from the fingerprint of the entanglement spectra and the full set of topologically degenerate ground states. We also show a coexisting stripe order in the CSL phase. The nature of the phase transitions is explored in Sec.~\ref{phasetransition}. Sec.~\ref{summary} contains a discussion and summary.

\section{Model and method}
\label{modelandmethod}

The spin-1 $J_{1}$-$J_{2}$-$J_\chi$ Heisenberg model on the square lattice is defined as
\begin{eqnarray}
\label{eq1}
H &=& J_{1}\sum\limits_{\left\langle i,j\right\rangle
}\mathbf{S}_{i}\cdot \mathbf{S}_{j}+J_{2}\sum\limits_{\left\langle \left\langle
i,j\right\rangle \right\rangle }\mathbf{S}_{i}\cdot \mathbf{S}_{j} \nonumber \\ 
&+& J_{\chi}\sum\limits_{i,j,k\in \triangle }\mathbf{S}_{i}\cdot (\mathbf{S}_{j}\times \mathbf{S}_{k}),
\end{eqnarray} 
where $\left\langle i,j\right\rangle$ and $\left\langle \left\langle i,j\right\rangle \right\rangle $ refer to the NN and NNN sites. $\left \{ i,j,k \right \}$ in the summation $\sum _{\Delta }$ refers to the three neighboring sites of the smallest triangle taken clockwise [see Fig.~\ref{Fig_phase_diagram}(a)]. 
Here we take $J_{1}=1$ as the energy unit.

We study the system using both the finite and infinite DMRG methods~\cite{white1992density,white1993density,schollwock2011density} with $U(1)$ spin symmetry on the cylinder geometry with circumference $L_{y} = 4 - 10$ lattice sites~\cite{ITensorandTenPy,itensor,tenpy}.
To determine the magnetic order, we directly compute the magnetic moments and also check the spin structure factor using the
DMRG calculation with spin $SU(2)$ symmetry~\cite{su2}, which gives consistent results.
In the infinite DMRG simulation, two or four columns are chosen as the unit cell to accommodate the magnetic orders. 
We keep up to M=10000 ($4000$) $U(1)$ [$SU(2)$] bond dimensions with a typical truncation error $\epsilon \sim 10^{-5}$. 
More benchmark results and details are shown in the Appendix~\ref{supp_convergence}.

\section{Phase diagram}
\label{phasediagram}

Through DMRG calculations, we establish a quantum phase diagram as shown in Fig.~\ref{Fig_phase_diagram}(c), which has three magnetic ordered phases.
For $J_{\chi } = 0$, we find a direct transition from the N\'eel to the stripe state at $J_2 \approx 0.546$~\cite{SSGong2018}. 
By turning on $J_\chi$, the $(\pi,\pi)$ N\'eel state and $(0,\pi)$ stripe state smoothly expand to finite-$J_\chi$ regime. 
For large $J_{\chi}$, we find a CSS with a large chiral order and the spin correlations consistent with the corresponding classical state~\cite{Trugman1995}. 
In the intermediate-$J_{\chi}$ regime sitting among these magnetic phases, we identify a CSL with the non-Abelian Pfaffian-type topological order~\cite{Moore1991,Greiter1991} by the quasi-degenerate patterns in the ES and three topologically degenerate ground states.
We also find that this topological state coexists with a stripe magnetic order, which we label as the CSL+stripe state.

For comparison, we depict a classical phase diagram of this model in Fig.~\ref{Fig_phase_diagram}(b), showing that the CSL emerges roughly around the transition lines between these magnetic order phases. 
This picture agrees with the guiding principle for searching for CSLs in spin-$1/2$ models~\cite{Hu2015,Wietek2015,Gong2015,gong2017global,huang2021quantum,hickey2017emergence}, suggesting that strong magnetic fluctuation still plays an important role in driving the CSLs even though their topological natures are different. 

\begin{figure}
\centering
\includegraphics[width=1\linewidth]{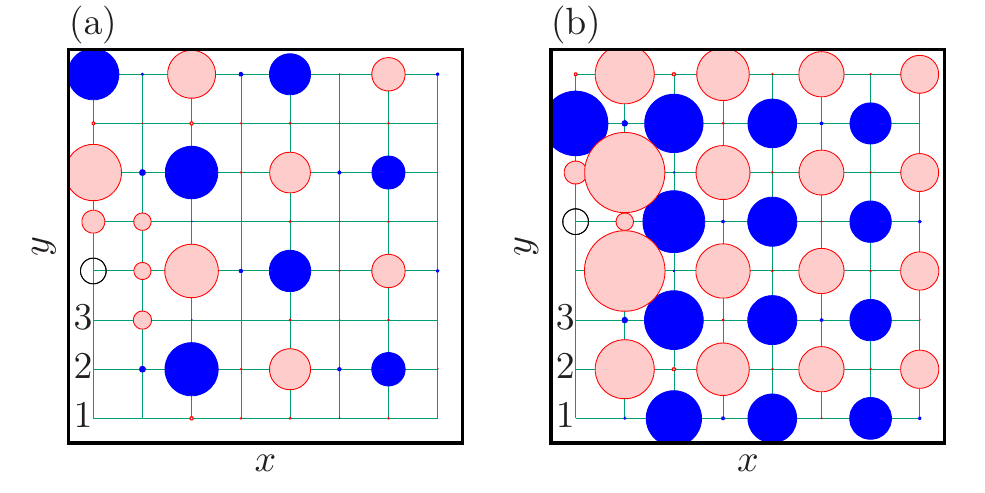}
\caption{\label{Fig_correlation_2D} Spin correlation function $\left \langle \mathbf{S}_{i} \cdot \mathbf{S}_{j} \right \rangle$ at $J_{2} = 0.5, J_{\chi} = 0.45$ in the CSS. (a) The black open circle is the reference site $i$, and all the spin correlations for the even reference sites have the same correlations. The blue solid (red shaded) circles represent positive (negative) spin correlation, and the radius represents the magnitude. (b) Spin correlations with the same parameters for the odd reference sites. The results are obtained with $L_{y} = 8$.}
\end{figure}

\section{Magnetic orders}
\label{magnetic}
We determine the magnetic orders using two methods, which give the same conclusions.   
First, we directly measure the magnetic moments $\langle S^{z} \rangle$ in the systems by applying a magnetic pinning field $h_z$ at the boundary of the cylinder~\cite{White2007}.
Second, we compute the spin structure factor 
without magnetic pinning fields using the $SU(2)$ DMRG and identify the magnetic order by the sharp Bragg peaks. 
With the help of a small pinning field, as shown in Figs.~\ref{Fig_phase_diagram}(d) and~\ref{Fig_phase_diagram}(e), $\langle S^{z} \rangle$ have large magnitudes in the bulk of the cylinder, with staggered and uniform signs for the N\'eel and stripe ordered states, respectively. 
In the insets, the spin structure factors also show the strong characteristic peaks supporting the magnetic orderings. 
For the CSL+stripe state in Fig.~\ref{Fig_phase_diagram}(f), $\langle S^{z} \rangle$ is also found to saturate at a finite value and the spin structure factor has the ($0, \pi$) peak characterizing a stripe order.

In the CSS shown in Fig.~\ref{Fig_phase_diagram}(g) with boundary pinning fields, $\langle S^{z} \rangle$ are large on every other site, agreeing with the classical spin configuration where the magnetic moments at other sites are lying down in the $xy$ plane~\cite{Trugman1995} (also see Appendix~\ref{supp_CSS}). 
The noncoplanar magnetic order of the state is confirmed by  a ``multi-$Q$'' feature in the spin structure factor, with two peaks at  $(0,\pi)$ and $(\pi,0)$ momenta and two satellite peaks at $(\pm\pi/2,\pm\pi/2)$, as shown in the inset in Fig.~\ref{Fig_phase_diagram}(g).
Alternatively, this feature can be viewed from the spin correlations in real space. As shown in Fig.~\ref{Fig_correlation_2D}(a) for the given reference site, the long-ranged spin correlations have a period of $4$ on the even sites, which corresponds to the $(\frac{\pi}{2}, \frac{\pi}{2})$ peak in the spin structure factor. The spin correlations on the odd sites shown in Fig.~\ref{Fig_correlation_2D}(b) have a period of $2$, which corresponds to the $(\pi,0)$ and $(0,\pi)$ peaks.

\begin{figure*}
\centering
\includegraphics[width=1\linewidth]{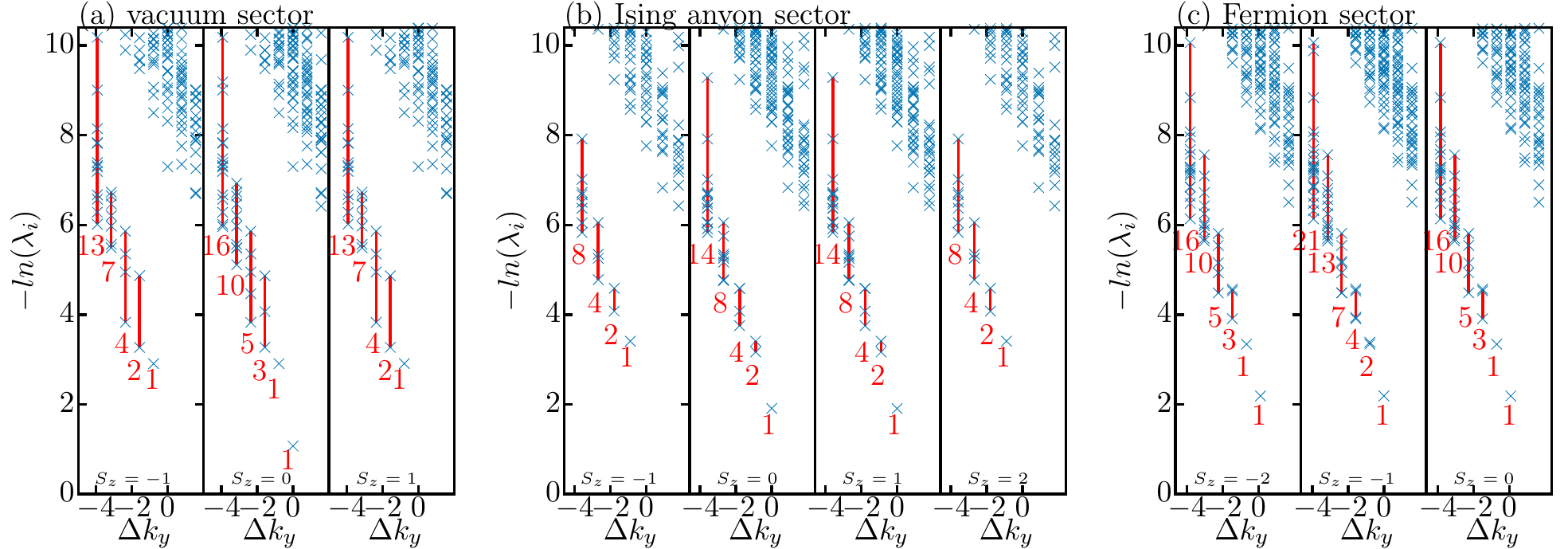}
\caption{Entanglement spectra of (a) the vacuum sector, (b) Ising anyon sector, and (c) fermion sector at $J_{2}=0.45,J_{\chi }=0.3$. The ESs in (a) and (c) are obtained with $L_{y}=8$, and the ES in (b) are obtained with $L_{y}=7$. $\lambda _{i}$ refers to the eigenvalues of the reduced density matrix of the half system, and $\Delta k_{y}= k_{y} - k_{y}^{0}$ is the relative momentum that has an increasement of $\frac{2\pi }{L_{y}}$ ($k_{y}^{0}$ is the momentum of the state with the largest eigenvalue in each sector). Each $S_{z}$ sector is separated using the conservation of total $S_{z}$. For a given $S_{z}$ and momentum, the quasi-degenerate eigenvalues are defined as the lower levels that are separated by the entanglement gap, which are labeled by the number below each momentum. 
}
\label{Fig_Spectrum_CSL}
\end{figure*}

\section{Coexisting non-Abelian CSL and stripe order}
\label{coexisting}
The ES extracted from the ground state provides a powerful tool to identify topological CSLs as it has a one-to-one correspondence to the edge spectra \cite{Haldane2008}. 
In the intermediate-$J_{\chi}$ regime, we find three nearly degenerate ground states using infinite DMRG simulation with randomly initialized states~\cite{Cincio2013}.
As shown in Fig.~\ref{Fig_Spectrum_CSL}, the quasi-degenerate patterns in the ESs are consistent with the tower of states of all three sectors described by the non-Abelian $SU(2)_{2}$ Wess-Zumino-Witten theory~\cite{ZLiu2012,WZhu2015}, which are identified as the vacuum sector, the Ising anyon sector, and the fermion sector.
In Fig.~\ref{Fig_Spectrum_CSL}(a), the leading ES of the vacuum sector shows a quasi-degenerate pattern of \{1,1,3,5,10,16,...\} in the $S_z = 0$ sector and \{1,2,4,7,13,...\} in the $S_{z} = \pm 1$ sectors (the higher degenerate levels are not observed due to the limited momentum numbers in finite-size systems).
The leading ES of the Ising anyon sector in Fig.~\ref{Fig_Spectrum_CSL}(b) shows the same pattern, \{1,2,4,8,...\}, for each given $S_{z}$ sector. 
In particular, the ES of the vacuum sector has a symmetry about $S_{z}=0$, while it is symmetric to $S_{z}=1/2$ in the Ising anyon sector, corresponding to a spin-$1/2$ quasiparticle created at the open edge.
As shown in Fig.~\ref{Fig_Spectrum_CSL}(c), the leading ES of the fermion sector has degenerate patterns similar to those in the vacuum sector except for the symmetry about $S_{z}=-1$, which indicates the created spin-1 excitation at the edge of the cylinder.

\begin{figure}
\centering
\includegraphics[width=1\linewidth]{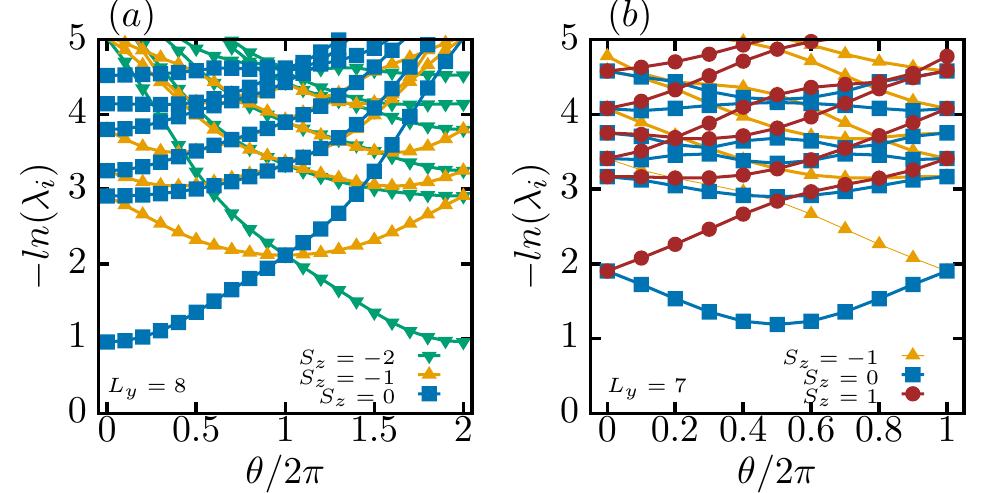}
\caption{ES flow with adiabatically inserted flux $\theta$ starting from (a) the vacuum sector and (b) the Ising anyon sector at $J_{2}=0.45,J_{\chi }=0.3$. The eigenvalues in different $S_{z}$ sectors are labeled by different symbols.}
\label{Fig_flux_insertion}
\end{figure}

\begin{figure}
\centering
\includegraphics[width=0.9\linewidth]{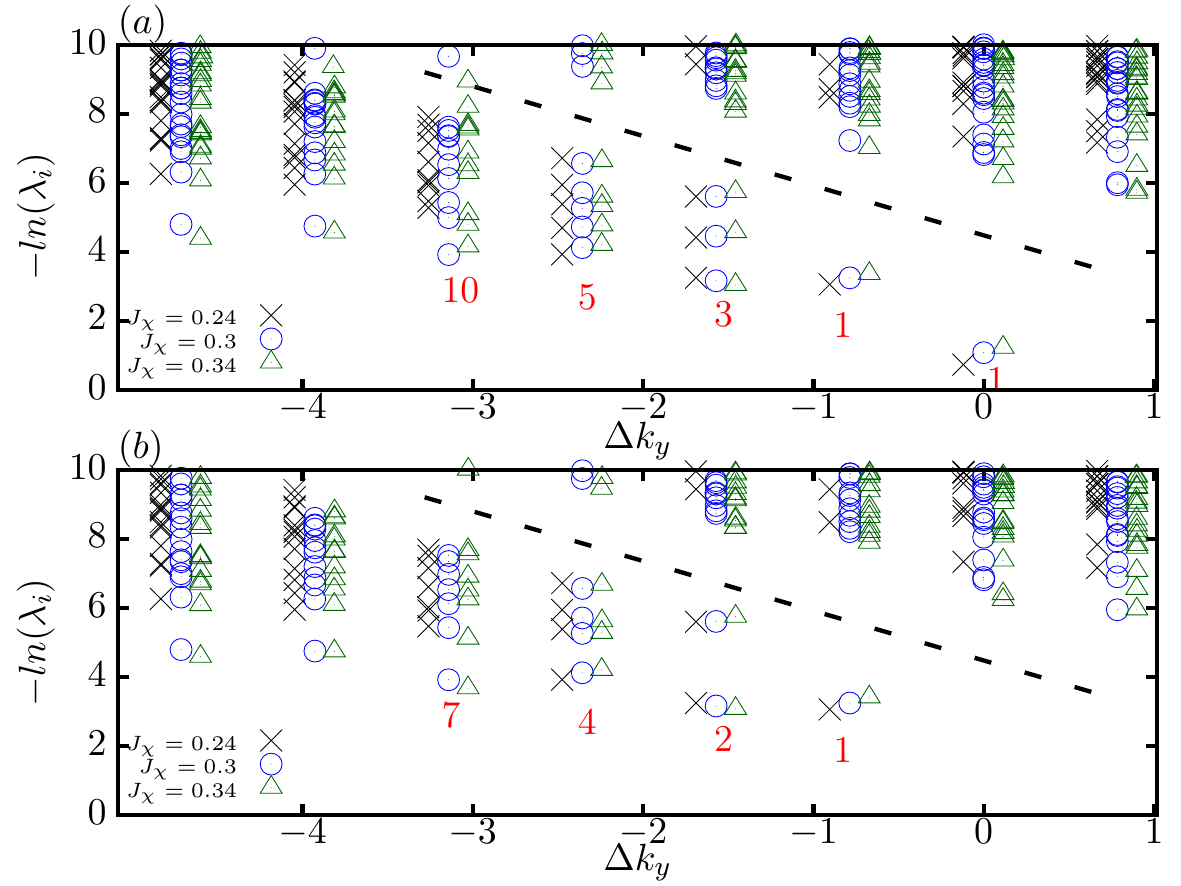}
\caption{\label{Fig_flux_Jx} The entanglement spectra for various $J_{\chi}$ at $J_{2}=0.5$, obtained with $L_{y}=8$ in the vacuum sector. (a) and (b) are the lowest and second-lowest parts of the entanglement spectra, respectively. A minor shift in $\Delta k_{y}$ is applied for different $J_{\chi}$ in order to view the quasi-degenerate eigenvalues, which are separated by the entanglement gap indicated by the dashed line.}
\end{figure}

Here we stress that the selection of each topological sector of this non-Abelian CSL is well controlled. 
First, we find that the two Abelian sectors are more stable on an even-width cylinder and the Ising anyon sector is energetically favored on an odd-width cylinder. 
This observation can be understood based on the generalized Pauli principle in the thin-torus limit of the bosonic Moore-Read Pfaffian state, i.e. no more than two particles in two consecutive orbitals~\cite{Bernevig2008,ZLiu2012,WZhu2015}. 
Thus, the Abelian vacuum and fermion sectors, related to the patterns $[20]$ and $[02]$ (period of $2$ orbitals), are favorable under even-width cylinders, while the non-Abelian Ising anyon sector corresponding to $[11]$ (period of $1$ orbital) survives on odd-width cylinders. 
Alternatively, the even-odd effect can also be derived based on the boundary condition of Majorana fermions: the odd-width cylinders leading to the anti-periodic boundary condition guarantee the existence of a Majorana zero mode~\cite{Sarma2015,ZXLiu2012}.
Second, the two Abelian sectors can be smoothly connected by adiabatically inserting flux in the cylinder. 
The flux $\theta$ adds a phase factor to the spin-flip terms $S_{i}^{+} S_{j}^{-}\rightarrow e^{i\theta }S_{i}^{+} S_{j}^{-}$ for $j \rightarrow i$ across the boundary from the top and their Hermitian conjugate terms.
As shown in Fig.~\ref{Fig_flux_insertion}(a), the ES of the vacuum sector at zero flux ($\theta=0$) is symmetric to $S_{z}=0$.
With growing $\theta$, the ES adiabatically evolves into the fermion sector at $\theta = 2\pi$ with a symmetry about $S_{z} = -1$, which 
indicates a net spin $\Delta S=1$ transferred from one edge of the cylinder to the other one. 
Furthermore, the ES evolves back to the vacuum sector with a symmetry about $S_{z}=-2$ at $\theta = 4\pi$, indicating the total transferred spin $\Delta S=2$. 
On the other hand, the Ising anyon sector evolves to itself by threading one flux quantum, as shown in Fig.~\ref{Fig_flux_insertion}(b). 
The evolution  of different sectors with flux insertion directly shows the gapless feature of the edge states~\cite{qi2012general,Avron2003}.
Furthermore, we show  the robustness of the vacuum sector topological ES in the CSL+stripe phase with fixed $J_2=0.5$ and varying $J_{\chi}$. As seen in Figs.~\ref{Fig_flux_Jx}(a) and~\ref{Fig_flux_Jx}(b), the quasidegenerate patterns remain robust in the momentum sectors associated with the lowest four groups of entanglement spectra levels.

To establish the coexisting stripe order, we also study the finite-size scaling of magnetic order parameters defined as $m^{2}(\mathbf{k})=\frac{1}{N^{2}}\sum_{i,j}\left \langle \mathbf{S}_{i} \cdot \mathbf{S}_{j} \right \rangle e^{i\mathbf{k}\cdot (\mathbf{r}_{i}-\mathbf{r}_{j})}$, where $i$ and $j$ are summed over the middle $N=L_{y} \times L_{y}$ sites. 
Figure~\ref{Peak_chiral_order}(a) shows the stripe order parameters at $(0, \pi)$, which are obtained with the $SU(2)$ DMRG with extrapolation to infinite bond dimensions (see more details in Appendix~\ref{supp_convergence}). 
The finite $m^{2}(0, \pi)$ in the two-dimensional (2D) limit pins down the stripe order in the CSL+stripe regime.

\section{Phase transition}
\label{phasetransition}
\begin{figure}
\centering
\includegraphics[width=1\linewidth]{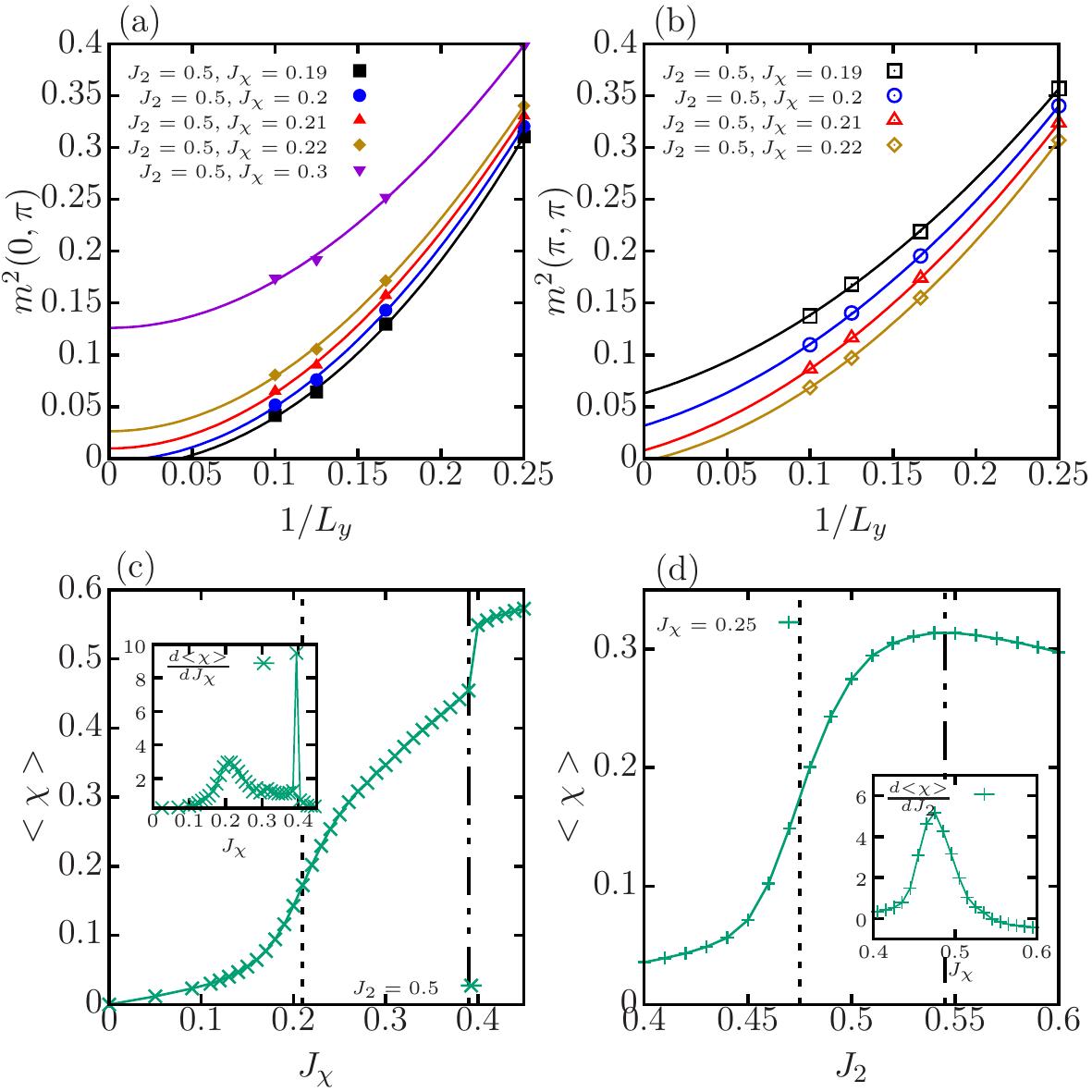}
\caption{Magnetic order parameters $m^{2}(\mathbf{k})$ at (a) $ (0, \pi)$ and (b) $ (\pi, \pi)$ are shown for various $J_{\chi}$~\cite{note} at $J_{2}=0.5$. The lines in (a) are polynomial fittings up to the second order of $1/N$. In (b) we find a more accurate fitting curve up to the second order of $1/\sqrt{N}$ close to the transition point. (c) is the chiral order $\langle \chi \rangle$ at $J_{2}=0.5$ for various $J_{\chi}$. The states from left to right are N\'eel, CSL+stripe, and CSS. The dashed lines are determined by the peaks in the derivative of $\langle \chi \rangle$, as shown in the inset. (d) is the chiral order $\langle \chi \rangle$ at $J_{\chi}=0.25$ for various $J_{2}$, showing the N\'eel, CSL+stripe, and stripe state from left to right. The left dashed line is determined by the peak in the derivative of $\langle \chi \rangle$ and the right dashed line is determined by the vanishing of quasi-degenerate patterns in the ESs (see Appendix~\ref{supp_entan}). The results in (c) and (d) are obtained with $L_{y}=8$.
}
\label{Peak_chiral_order}
\end{figure}

We first detect phase transitions using magnetic order parameters. 
We show $m^{2}(\pi, \pi)$ and $m^{2}(0, \pi)$ from the N\'eel to the CSL+stripe phase at $J_{2}=0.5$ in Figs.~\ref{Peak_chiral_order}(a) and~\ref{Peak_chiral_order}(b).
After finite-size scaling, the N\'eel order $m^{2}(\pi, \pi)$ vanishes at $J_{\chi} \approx 0.21$ and the stripe order $m^{2}(0, \pi)$ starts to develop at the same $J_{\chi}$, indicating a direct transition between the two phases.
As shown in Fig.~\ref{Peak_chiral_order}(c), the chiral order, defined as $\langle \chi \rangle = \frac{1}{4N}\sum\limits_{i,j,k\in \triangle }\langle \mathbf{S}_{i}\cdot (\mathbf{S}_{j}\times \mathbf{S}_{k}) \rangle$, also rises quickly at $J_{\chi} \approx 0.21$ with a peak in its first-order derivative of $J_{\chi}$, suggesting a continuous transition to the CSL+stripe. 
The same transition can also be found by varying $J_{2}$, as shown in Fig.~\ref{Peak_chiral_order}(d) for $J_{\chi} = 0.25$.
However, the nature of the transition around $J_2=0.45, J_{\chi}=0.3$ becomes more complicated. 
We identify possible two-step transitions as the stripe order disappears first inside the shaded region while the topological order remains robust and another transition to the N\'eel state happens at lower $J_{\chi}$. We leave this intriguing possibility to future study.

With further increasing $J_{\chi}$ at $J_{2}=0.5$, $m^{2}(\frac{\pi}{2}, \frac{\pi}{2})$ shows a sudden rise at $J_{\chi } \approx 0.4$ (see details in Appendix~\ref{supp_CSS}), and the chiral order has a sharp jump, as shown in Fig.~\ref{Peak_chiral_order}(c), indicating a first-order transition from the CSL+stripe state to the CSS.
To determine the boundary between the CSL+stripe and stripe states, we identify the melting of the low-lying quasi-degenerate patterns in the ESs as a probe of vanished topological order~\cite{SuppMaterial}, which determines a topological quantum phase transition.

\section{Summary and discussion}
\label{summary}
We have identified a non-Abelian Pfaffian-type CSL with coexisting stripe magnetic order in an extended spin-1 quantum antiferromagnet on the square lattice using the unbiased DMRG calculation. 
We established the topological nature using the characteristic entanglement spectra and three-fold topological degeneracy. 
Notice that the previously found CSLs in spin-$1/2$ systems are all non magnetic~\cite{Bauer2014,He2014,Gong2014,Messio2012,Hu2015,Kumar2014,Wietek2015,Gong2015,He2015c}, which indicates that the reduced quantum fluctuations in spin-1 system may leave a room for promoting the coexistence ofmagnetic order with a QSL.
Our findings not only demonstrate an explicit example with the coexistence of fractionalized excitations and magnetic order in frustrated quantum magnets, but also inspire a new search for exotic phases in other interesting systems, including the moir\'e superlattices, where effective spin models with $SU(4)$ symmetry may lead to new QSLs~\cite{Wu2018,Regan2020,Tang2020}.

Finally, we give some remarks on the possible experimental realization of the system. 
Recent studies on FeSe have shown a paramagnetic parent state with both N\'eel and stripe spin fluctuations~\cite{wang2016magnetic} and a pressure-induced stripe magnetic phase~\cite{Wang2016pressure}, which can be considered an effective spin-1 system~\cite{gong2017possible}. 
The chiral interaction can be induced by the orbital coupling of applied magnetic field to the underlying electrons~\cite{Sen1995}. 
For conventional spin systems with time-reversal symmetry, a dynamical spin chiral term can also be generated by the circularly polarized light~\cite{claassen2017dynamical}. 
\\

\begin{acknowledgments}
W.Z. thanks X. Y. Dong, Z. X. Liu and X. L. Wang for simulating discussions. 
This work was supported by the U.S. Department of Energy, Office of Science, Advanced Scientific Computing Research and Basic Energy Sciences, Materials Sciences and Engineering Division, Scientific Discovery through Advanced Computing (SciDAC) program under the grant number DE-AC02-76SF00515 (Y.H., H.C.J., D.N.S.). W.Z. was supported by National Science Foundation of China under project number 92165102. S.S.G. was supported by the NSFC grants No. 11874078 and 11834014.
\end{acknowledgments}

\newcommand{\beginsupplement}{%
        \setcounter{table}{0}
        \renewcommand{\thetable}{S\arabic{table}}%
        \setcounter{figure}{0}
        \renewcommand{\thefigure}{S\arabic{figure}}%
        \setcounter{equation}{0}
        \renewcommand{\theequation}{S\arabic{equation}}%
        }

\appendix
\beginsupplement
\section{Convergence of the numerical results}
\label{supp_convergence}

\begin{figure}
\centering
\includegraphics[width=1\linewidth]{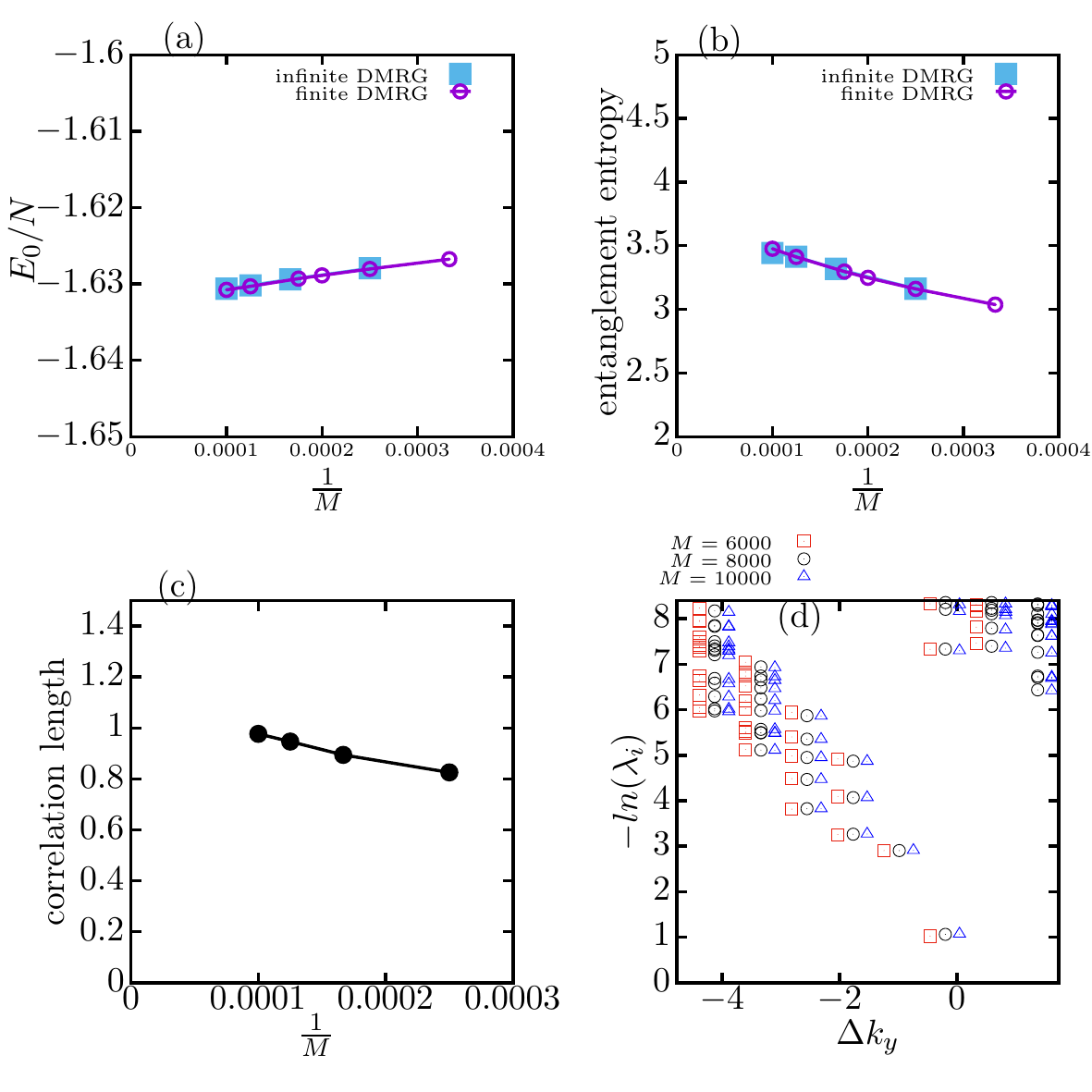}
\caption{\label{FigS:convergence} DMRG bond dimension dependence of (a) the ground-state energy per site, (b) the entanglement entropy, (c) the correlation length, and (d) the entanglement spectra. The finite DMRG results are calculated with $L_{x}=24, L_{y}=8$, and the infinite DMRG results are calculated with $L_{y}=8$. The ground-state energy is averaged over the half of the sites in the middle for the finite DMRG results in order to minimize the boundary effect. All the results are obtained at $J_{2}=0.45, J_{\chi}=0.3$.}
\end{figure}

\begin{figure}
\centering
\includegraphics[width=1\linewidth]{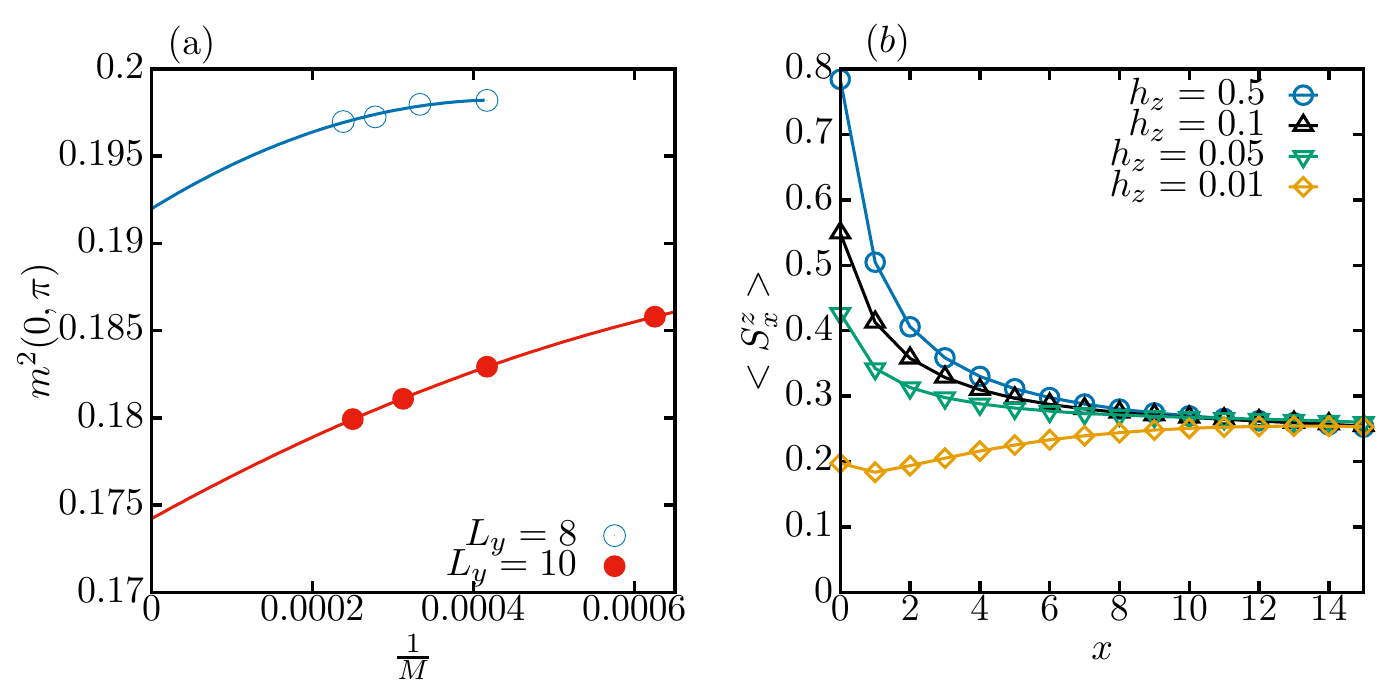}
\caption{\label{FigS:M_hz_order} Details of the DMRG calculation to determine magnetic order. (a) is the $SU(2)$ bond dimension dependence of the stripe order parameter at $J_{2}=0.5,J_{\chi}=0.3$. For $L_{y}=4$ and $6$ the order parameter remains almost unchanged with increasing bond dimension $M$, so the extrapolations are not shown here. (b) is the magnetic moment $\langle S^{z} \rangle$ with distance away from the edge for various pinning fields $h_{z}$ at $J_{2}=0.5,J_{\chi}=0.32$, obtained with $L_{y}= 8$.
}
\end{figure}

In order to check the numerical convergence, we obtain the density matrix renormalization group (DMRG) results by increasing the bond dimension $M$.  
As an example, we show the results for $J_2 = 0.45, J_{\chi} = 0.3$ in Fig.~\ref{FigS:convergence}, including the ground-state energy, entanglement entropy, and entanglement spectra, which all change slightly with increasing bond dimension and thus demonstrate the good convergence of our DMRG results.
Also, we can find highly consistent results from the finite and infinite DMRG simulations in Figs.~\ref{FigS:convergence}(a) and~\ref{FigS:convergence}(b).
Furthermore, we compute the correlation length to estimate the finite-size effect in our results.
The correlation length is defined as $1 / \ln(\epsilon_{1} / \epsilon_{2})$, where $\epsilon_{1}$ ($\epsilon_{2}$) is the largest (second largest) eigenvalue of the transfer matrix. 
In Fig.~\ref{FigS:convergence}(c) we show that the correlation length can be extrapolated to around $1.2$ in the limit of infinite bond dimensions, which is much smaller than the cylinder widths in our calculation. 
Thus, our DMRG results, such as the entanglement spectra, can provide reliable and accurate approximations of the physics in the 2D thermodynamic limit.

\subsection{Details of determining the magnetic order}

The stripe order parameter $m^{2}(0, \pi)$ defined in the main text has been shown to be extrapolated to a finite value in the chiral spin liquid (CSL) + stripe state. 
We calculate $m^{2}(0, \pi)$ by keeping different $SU(2)$ bond dimensions $M$ for $L_{y} = 8,10$. 
As shown in Fig.~\ref{FigS:M_hz_order}(a), the order parameters $m^{2}(0, \pi)$ slightly decay with increasing $M$ and remain finite in the infinite-$M$ limit. We use the extrapolated data in the main text.

In the calculation of magnetic moments $\langle S^{z} \rangle$, we have also tested the results by using various pinning fields $h_{z}$. 
As shown in Fig.~\ref{FigS:M_hz_order}(b), $\langle S^{z} \rangle$ remain almost the same in the bulk of the system for different values of $h_{z}$, indicating the robust magnetic moments.

\begin{figure*}
\centering
\includegraphics[width=1\linewidth]{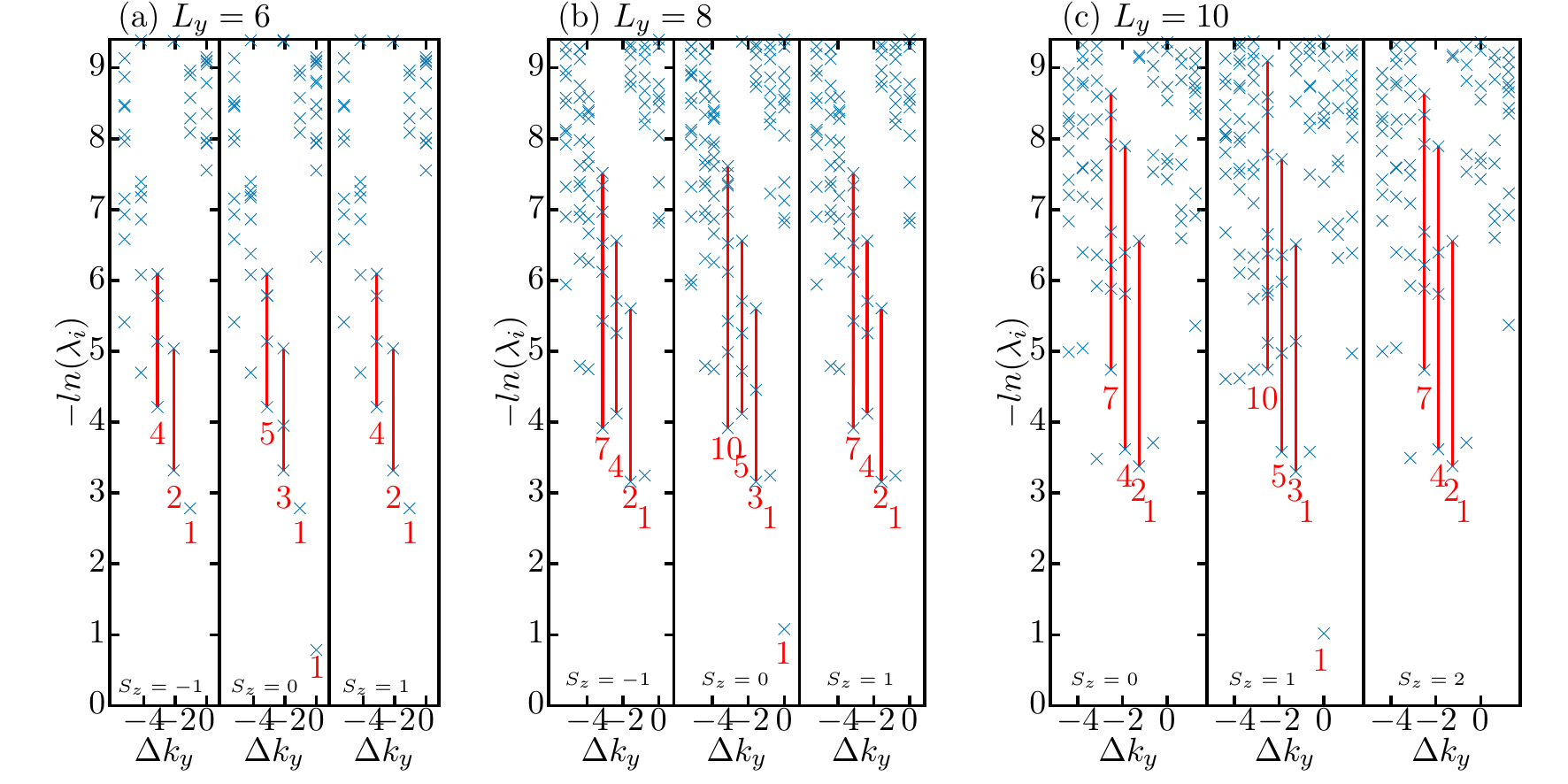}
\caption{\label{spectrum_CSL2} The entanglement spectra at $J_{2}=0.5, J_{\chi}=0.3$ for (a) $L_y = 6$, (b) $L_y = 8$, and (c) $L_y = 10$ in the CSL+stripe regime. Only the vacuum sector is shown.}
\end{figure*}

\begin{figure*}
\centering
\includegraphics[width=1\linewidth]{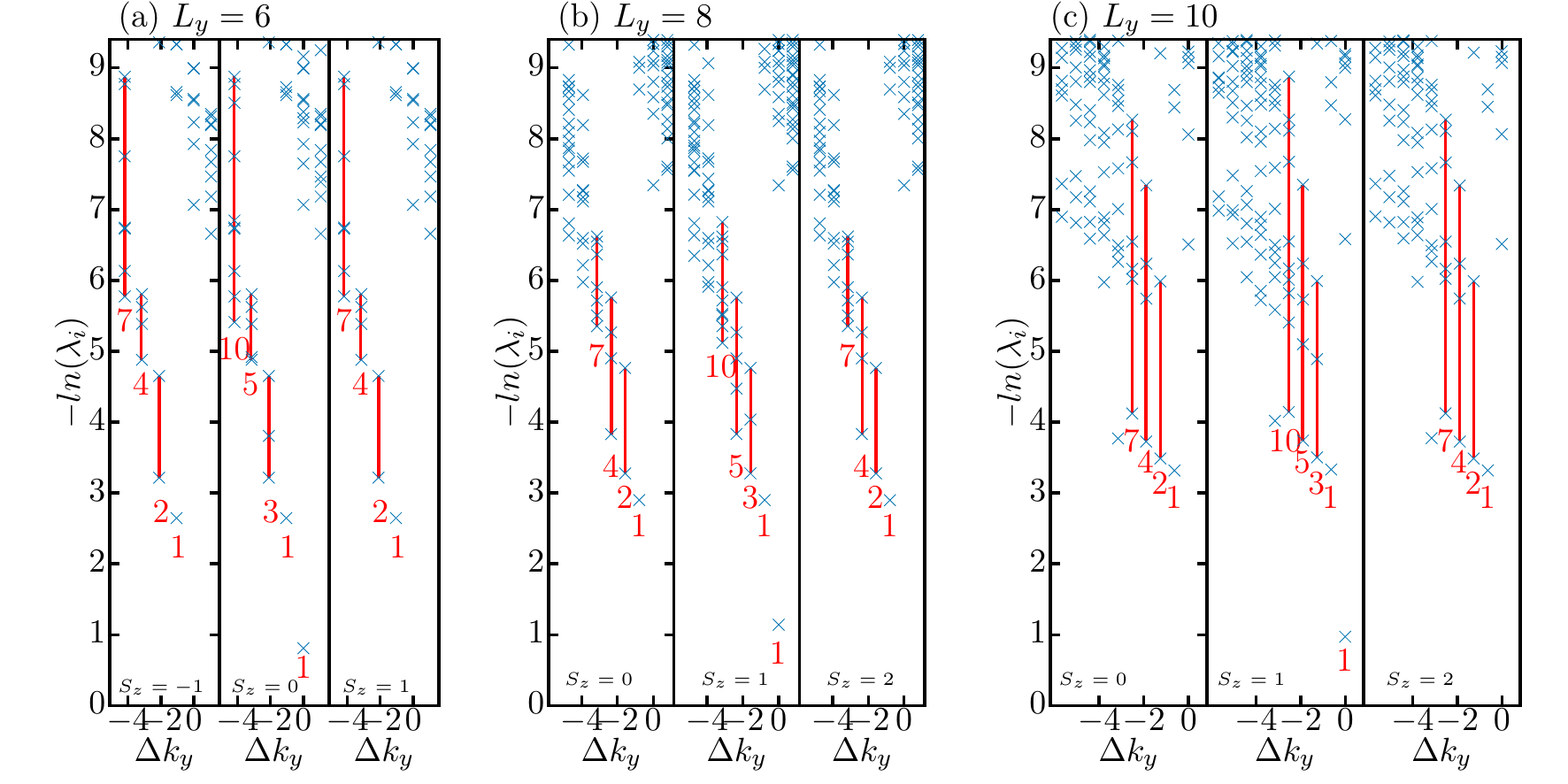}
\caption{\label{spectrum_Neel_CSL} The entanglement spectra at $J_{2}=0.42, J_{\chi}=0.34$ for (a) $L_y = 6$, (b) $L_y = 8$, and (c) $L_y = 10$ in the shaded regime. Only the vacuum sector is shown.}
\end{figure*}

\begin{figure*}
\centering
\includegraphics[width=0.95\linewidth]{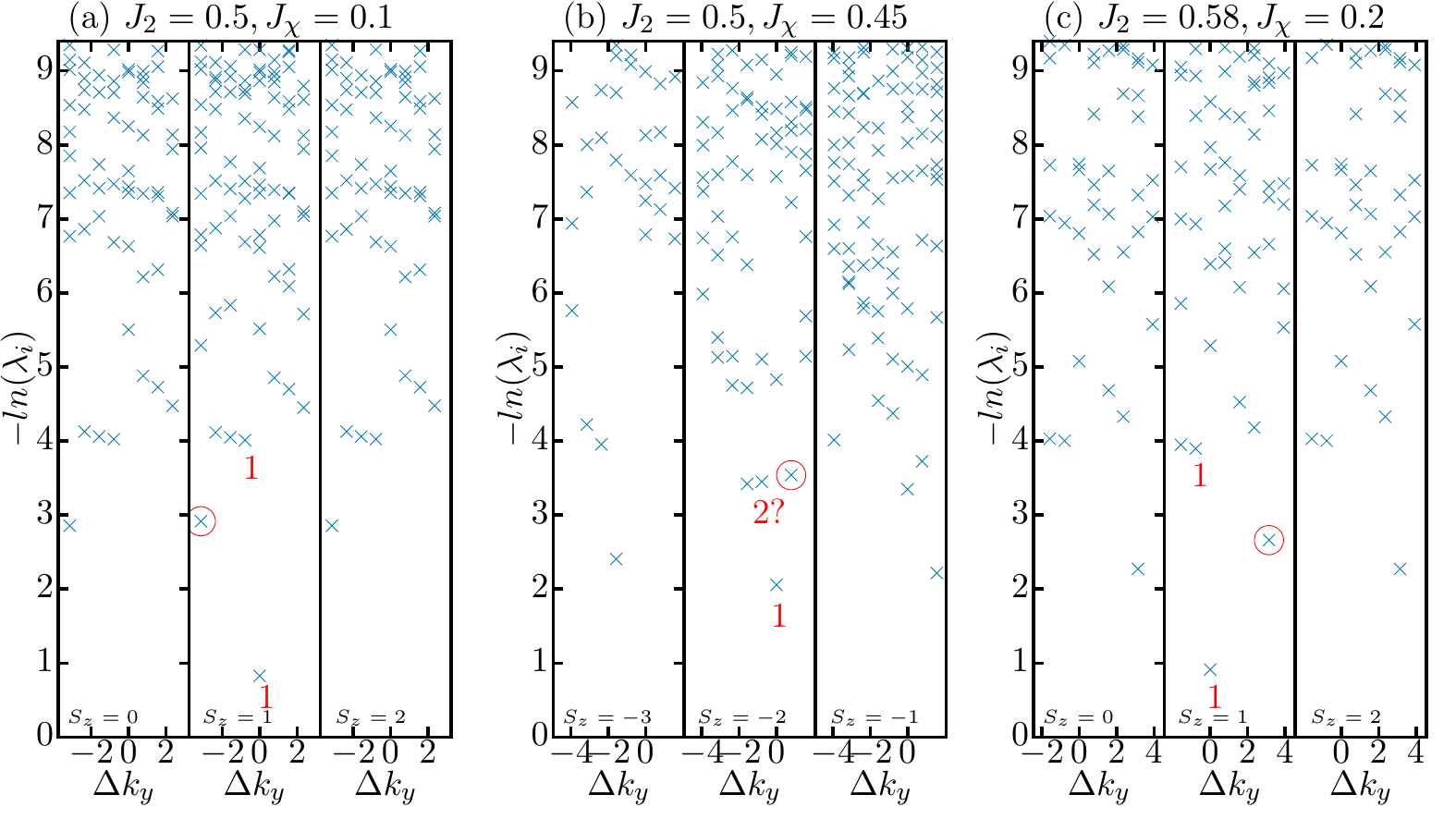}
\caption{\label{Neel_CSS} The entanglement spectra at (a) $J_{2}=0.5, J_{\chi}=0.1$ (N\'eel state), (b) $J_{2}=0.5, J_{\chi}=0.45$ (CSS), and (c) $J_{2}=0.58, J_{\chi}=0.2$ (stripe state) obtained with $L_{y}=8$. There is no counting of the quasi-degenerate eigenvalues.}
\end{figure*}

\section{Entanglement spectra}
\label{supp_entan}

\begin{figure*}
\centering
\includegraphics[width=1\linewidth]{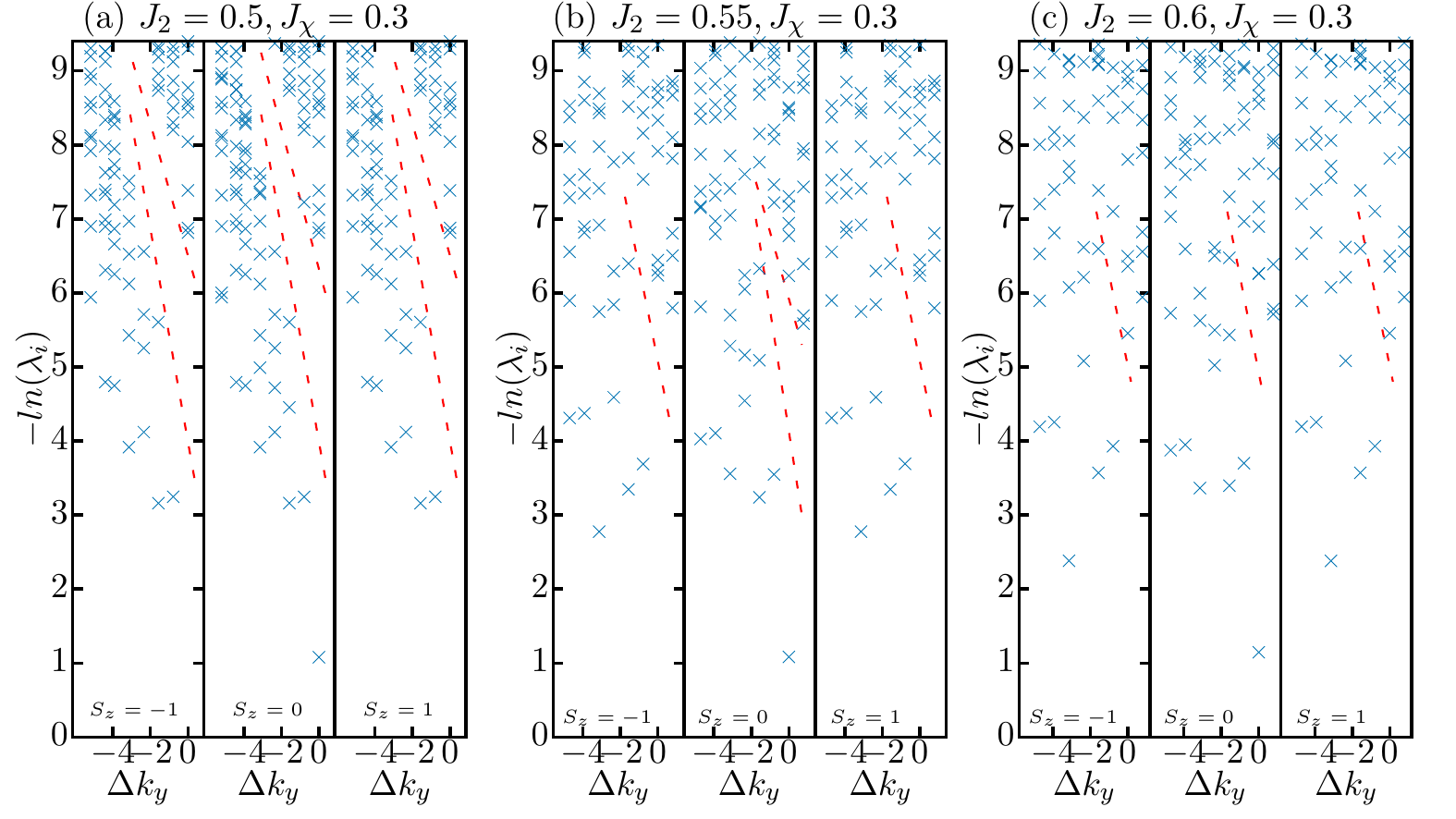}
\caption{\label{entan_spec_J2_boundary} The entanglement spectra at $J_{\chi}=0.3$ for various $J_{2}$, obtained with $L_{y}=8$. The red dashed lines are guides to the eye.}
\end{figure*}

In Figs.~\ref{spectrum_CSL2} and~\ref{spectrum_Neel_CSL}, we show the quasi-degenerate patterns of the entanglement spectra for more parameter points in both the CSL+stripe regime and the shaded regime.
We can find robust quasidegenerate patterns for various lattice sizes, which characterize the topological nature of the non-Abelian CSL. 
In comparison, the entanglement spectra have no such quasi-degenerate pattern in the N\'eel, stripe, and chiral spin states as shown in Fig.~\ref{Neel_CSS}.


The quasidegenerate patterns vanish as $J_{2}$ increases from $0.5$ in the intermediate $J_{\chi}$, which indicates a phase transition from the CSL+stripe to stripe state. As shown in Fig.~\ref{entan_spec_J2_boundary}(a), the quasidegenerate eigenvalues in the CSL+stripe state are separated by a relatively large entanglement gap in every spin sector. In Fig.~\ref{entan_spec_J2_boundary}(b) we can still identify the quasi-degenerate eigenvalues in the $S_{z} = 0$ sector near the phase boundary, but additional low-lying eigenvalues have already mixed in the $S_{z} = -1$ and $1$ sectors. When the system enters the stripe state, the quasidegenerate patterns disappear, as shown in Fig.~\ref{entan_spec_J2_boundary}(c).

\subsection{Spin gap in the CSL+stripe regime}

\begin{figure}
\centering
\includegraphics[width=0.98\linewidth]{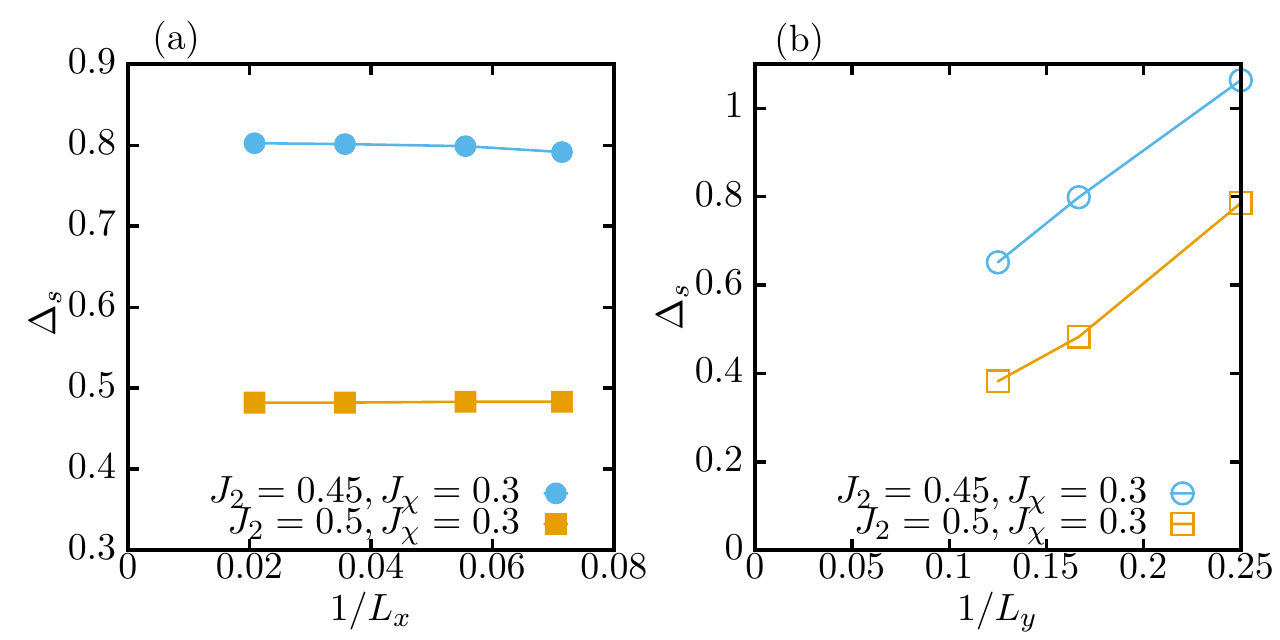}
\caption{\label{Spin_gap} Finite-size scaling of the spin gap in the CSL+stripe phase. (a) System length ($L_x$) dependence of the spin gap for $L_y = 6$. (b) System circumference ($L_y$) dependence of the spin gap obtained for cylinders with $L_{x} = 3 \times L_{y}$.}
\end{figure}

The spin gap can be obtained from the difference between the lowest energies of the total $S_{z}=0$ and $S_z = 1$ sectors. To avoid the edge excitations, we compute the spin excitations in the bulk. Here we compare two parameters in the CSL+stripe regime with the same $J_{\chi} = 0.3$. 
First, we study the system length ($L_x$) dependence of the spin gap for the given system circumference ($L_y$).
As shown in Fig.~\ref{Spin_gap}(a) for $L_y = 6$, the gaps are almost independent of the system length $L_x$.
By examining the spin correlations of the excited state, we can find that the excitations are mainly contributed by the spin flip along the circumference direction since the ground state has the $(0, \pi)$ order configuration, which is also consistent with our DMRG observation in Fig.~\ref{Spin_gap}(a).

Next, we study the scaling of the gap with $L_y$, as shown in Fig.~\ref{Spin_gap}(b). The gap near the boundary between the N\'eel and CSL+stripe state ($J_{2}=0.45$) is relatively large, because the magnetic order is weak and the gapless Goldstone mode may not be fully developed yet. 
In the center of the CSL+stripe state ($J_{2}=0.5$), the spin gap becomes smaller due to the developed stripe order and the associated gapless Goldstone mode. 
With growing system circumference, the spin gap decreases rapidly and may be appropriately scaled to vanishing, which indicates that the spin gap in the CSL+stripe state becomes smaller for larger systems and thus is consistent with gapless spin excitations in the CSL+stripe state.

\section{Numerical results for the chiral spin state}
\label{supp_CSS}

\subsection{Classical spin configurations}

\begin{figure}
\centering
\includegraphics[width=0.8\linewidth]{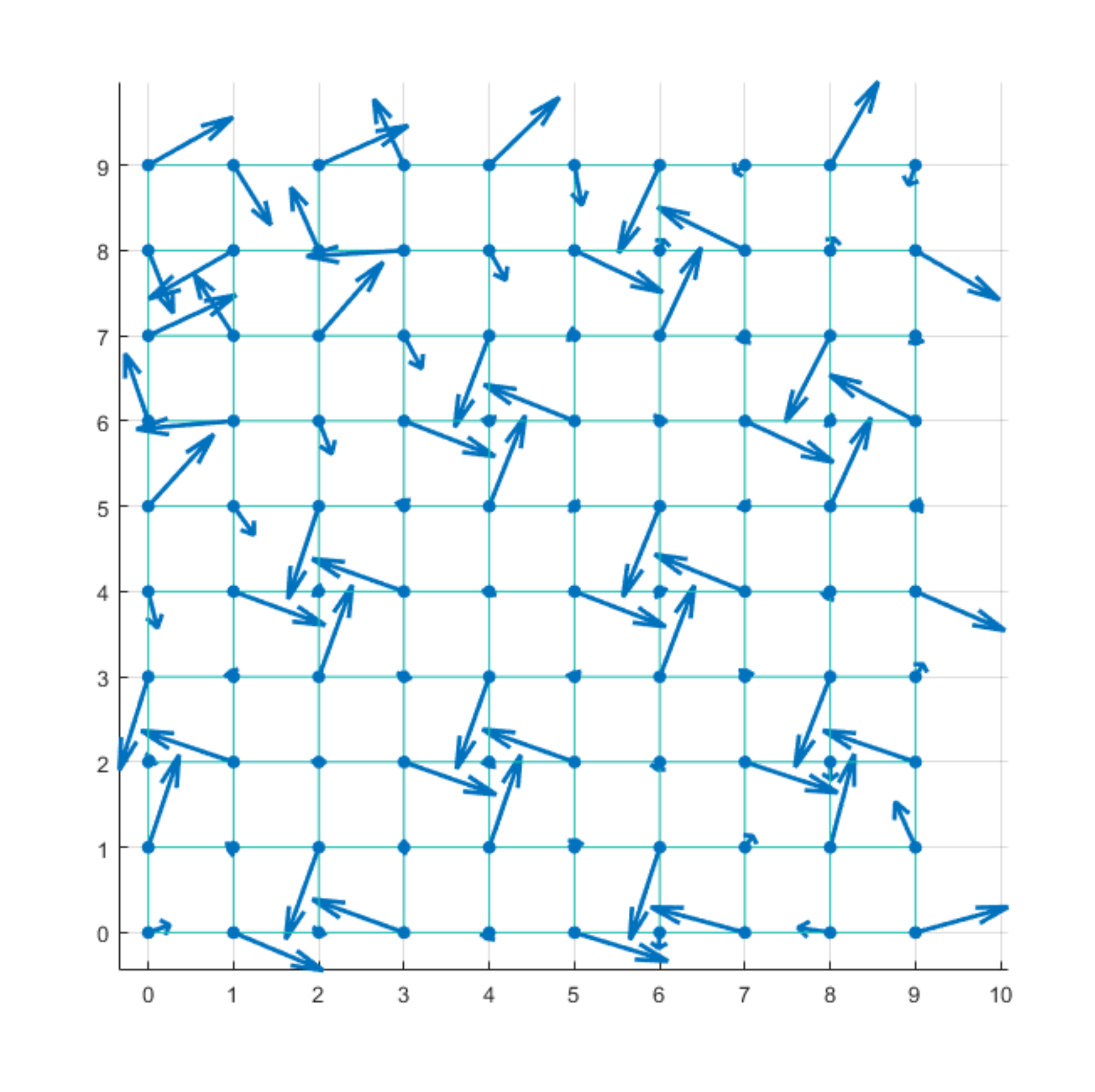}
\includegraphics[width=0.85\linewidth]{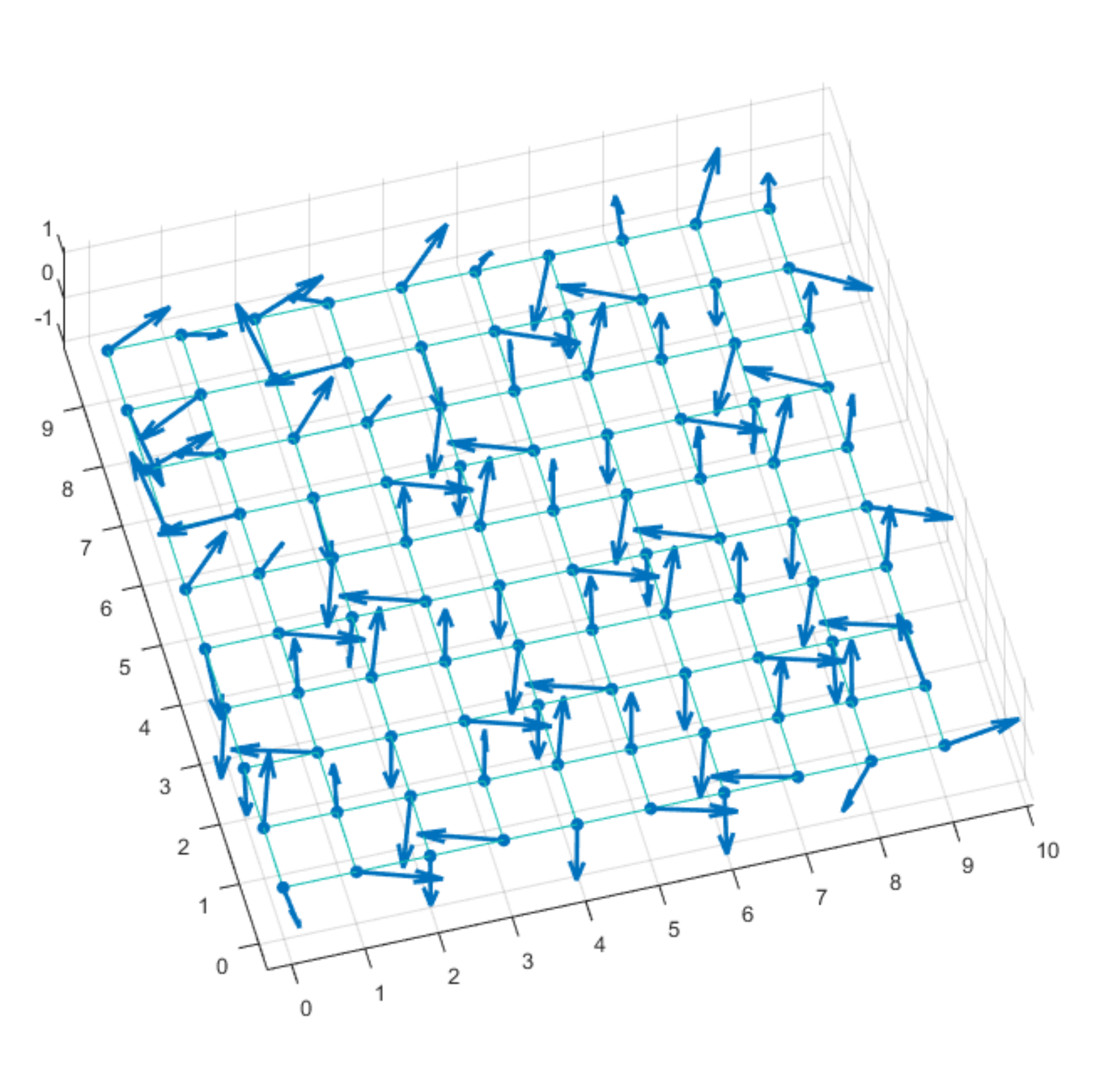}
\caption{\label{Classical_CSS} The spin configurations of the classical CSS using the classical Monte Carlo simulations on a $10 \times 10$ lattice, viewed from the top (the top panel) and side (the bottom panel).}
\end{figure}


The spin correlations in the CSS have configurations similar to the classical CSS, which is shown in Fig.~\ref{Classical_CSS}. While the spin pointing in the $xy$ plane has a period of $4$, the spin pointing in the $z$ direction has a period of $2$.

\subsection{Phase transitions}

\begin{figure}
\centering
\includegraphics[width=1\linewidth]{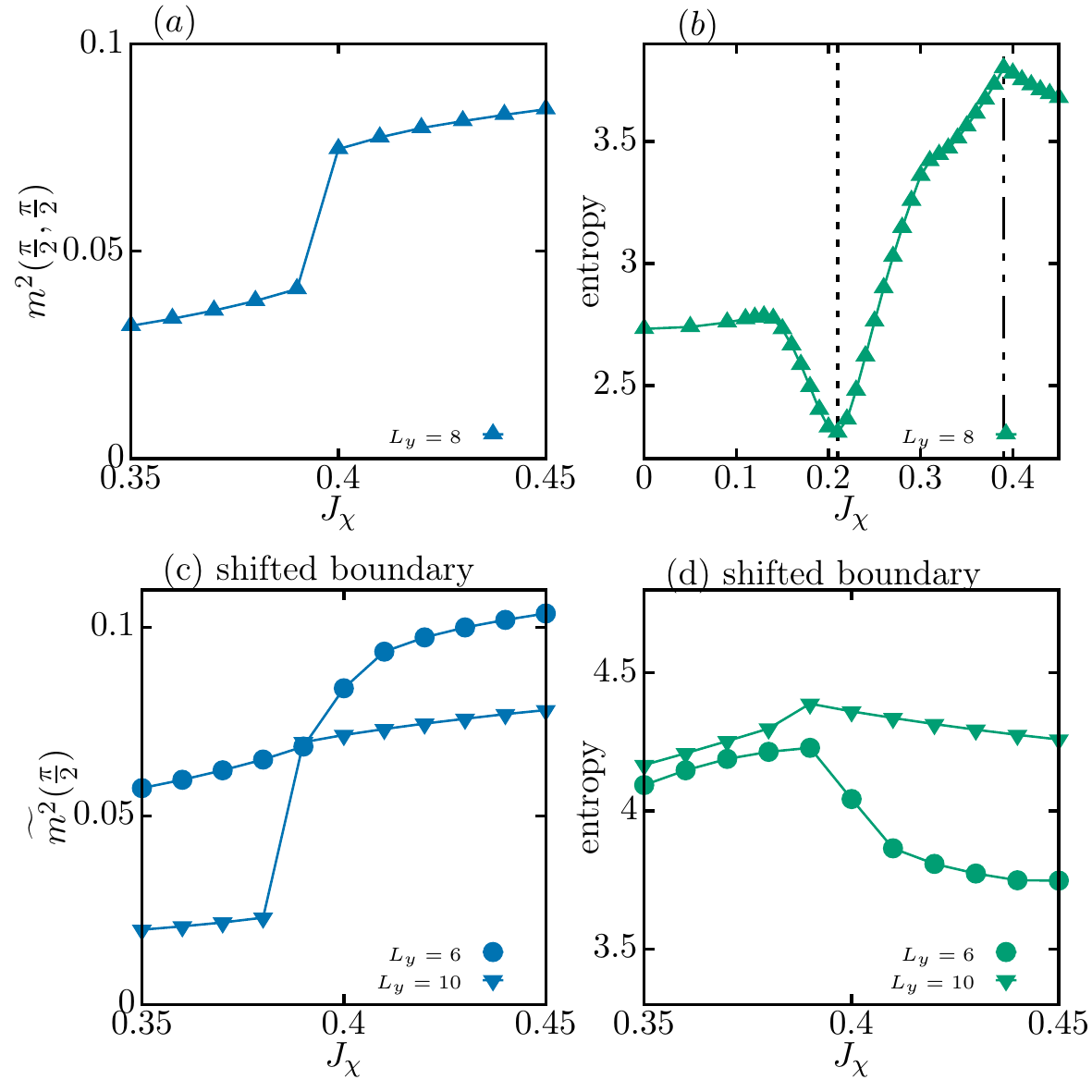}
\caption{\label{Fig_peak_entropy} Quantum phase transition from the CSL+stripe state to the CSS. (a) and (c) are different magnetic orders. (b) and (d) are the entanglement entropy versus $J_{\chi}$ near the phase boundary. (a) and (b) are obtained under the normal periodic boundary condition while (c) and (d) are obtained with the shifted boundary condition in order to be compatible with both magnetic orders. (b) shows the entanglement entropy in a larger $J_{\chi}$ regime, where the phase transitions from the N\'eel state to the  CSL+stripe state and from CSL+stripe state to the CSS are indicated by the two dashed lines. The left and right dashed lines are determined by the local minimum and maximum, respectively. All the results are obtained at $J_{2}=0.5$.}
\end{figure}

The phase transition from the CSL+stripe state to the CSS can be determined by the emergence of a finite $m^{2}(\frac{\pi}{2}, \frac{\pi}{2})$ order, because the two phases share the same structure factor peak at $(0, \pi )$. For $L_{y}=6,10$ we use the shifted boundary condition to be compatible with both phases. As a result, $k_{y}$ is no longer a conserved quantity and the lattice is rearranged into a two-leg ladder. Each ladder is constructed by connecting each pair of columns in the $y$ direction, and the modified order parameter becomes $\widetilde{m}^{2}(k)=\frac{1}{N^{2}}\sum_{i,j,Z}\left \langle \mathbf{S}_{i,Z} \cdot \mathbf{S}_{j,Z} \right \rangle e^{ik(r_{i,Z}-r_{j,Z})}$, where $Z=0,1$ is the leg index. As shown in Fig.~\ref{Fig_peak_entropy}(a), $m^{2}(\frac{\pi}{2}, \frac{\pi}{2})$ shows a sudden increase at $J_{\chi}\approx 0.4$ for $L_{y}=8$ [the finite $m^{2}(\frac{\pi}{2}, \frac{\pi}{2})$ below $J_{\chi }= 0.4$ results from the self-correlation contribution of $\langle \mathbf{S}_{i} \cdot \mathbf{S}_{i} \rangle = 2$ and thus is not a peak]. In Fig.~\ref{Fig_peak_entropy}(c) $\widetilde{m}^{2}(\frac{\pi}{2})$ also shows a sudden jump around the same $J_{\chi }$ for $L_{y}=6,10$, indicating a first-order transition that is consistent for different lattice sizes. 

In addition, we obtain the entanglement entropy near the phase transitions. A maximum value of the entanglement entropy can be found near the CSL+stripe to CSS phase transition for various $L_{y}$. As shown in Figs.~\ref{Fig_peak_entropy}(b) and~\ref{Fig_peak_entropy}(d), the entanglement entropy reaches its maximum at $J_{\chi }\approx 0.39$ for $L_{y}=6,8,10$, which is consistent with the phase transition.


\bibliography{spin1}

\begin{thebibliography}{74}
\bibitem [{not()}]{note}%
  \BibitemOpen
  \href@noop {} {}\bibinfo {howpublished} {The results at $J_{\chi}=0.3$ are
  obtained with $SU(2)$ DMRG because $U(1)$ calculations result in finite
  magnetic moments $<S^{z}>$. Results at other $J_{\chi}$ are obtained with
  $U(1)$ DMRG.}\BibitemShut {Stop}%
\end{thebibliography}

\end{document}